\documentclass[twoside]{ilcws10}
\usepackage[latin1]{inputenc}
\usepackage[dvips]{graphicx,epsfig,color}
\usepackage{wrapfig,rotating}
\usepackage{amssymb,amsmath,array}

\begin{document}
\title{
    Using Single Photons for WIMP Searches at the ILC
} 
\author{
  Koichi Murase$^1$,
  Tomohiko Tanabe$^2$,
  Taikan Suehara$^2$,
  Satoru Yamashita$^2$
  and Sachio Komamiya$^{1,2}$
\vspace{.3cm}\\
1- Department of Physics, Graduate School of Science, The University of Tokyo \\
7-3-1 Hongo, Bunkyo-ku, Tokyo 113-0033 - Japan \\
\vspace{.1cm}\\
2- International Center for Elementary Particle Physics (ICEPP), The University of Tokyo \\
7-3-1 Hongo, Bunkyo-ku, Tokyo 113-0033 - Japan \\
}

\maketitle

\begin{abstract}
We consider analysis targets at the International Linear Collider
in which only a single photon can be observed.
For such processes, we have developed
a method which uses likelihood distributions
using the full event information (photon energy and angle).
The method was applied to a search for neutralino pair production
with a photon from initial state radiation (ISR)
in the case of supergravity in which the
neutralino is the lightest supersymmetric particle.
We determine the cross section required to observe the
neutralino pair production with ISR as a function of the
neutralino mass in the range of 100 to 250~GeV.
\end{abstract}

\section{Introduction}
Weakly interacting massive particles (WIMPs) 
are important study targets in current and future collider experiments
since the WIMPs provide a solution to the dark matter problem.
In this study, we investigate the prospects of discovering
the direct pair production of WIMPs
at the International Linear Collider (ILC)
using the energy and angle spectra
of the photons from initial state radiation (ISR)~\cite{Presentation}.
We make the assumption that, among the particles beyond the Standard Model,
the WIMPs are the only kinematically accessible
at the center-of-mass (CM) energy of 500~GeV proposed for the ILC.
Similar past searches using single photons have been performed by e.g.~\cite{Lopez:1996ey,Akrawy:1990gy}.

\begin{figure}[hbt!p]
\centering
\begin{minipage}[t]{0.38\columnwidth}
\centering
\includegraphics[width=0.7\columnwidth]{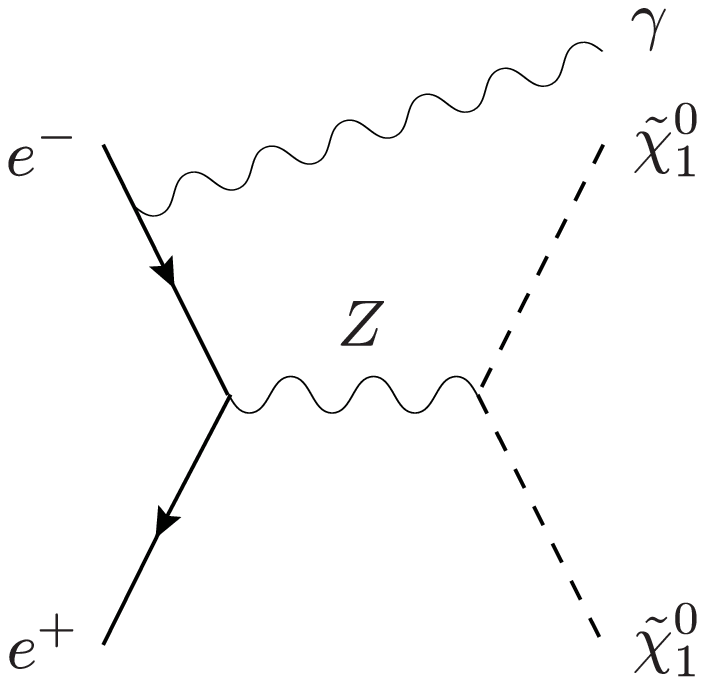}
\end{minipage}
\begin{minipage}[t]{0.38\columnwidth}
\centering
\includegraphics[width=0.7\columnwidth]{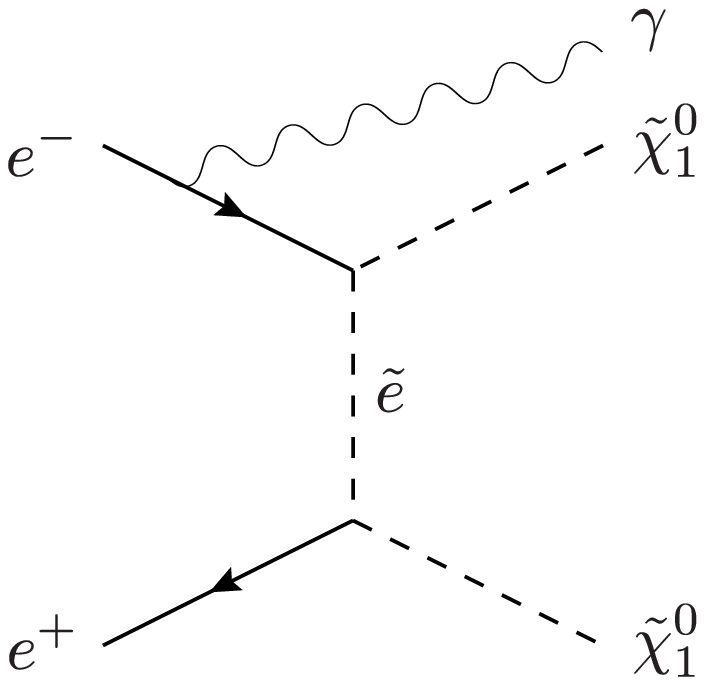}
\end{minipage}
\caption{
  Tree-level diagrams for neutralino pair production
  with initial state radiation:
  $s$-channel (left) and $t$-channel (right).
}
\label{fig:tchannel}
\end{figure}

Prospects of such measurements have been discussed 
in many literature~\cite{Fayet,Fargion:1995qb,Birkedal:2004xn,Konar:2009ae} and,
in particular, at the ILC~\cite{Bartels} which uses a cut-and-count approach.
Our method uses the two-dimensional energy and angle distributions of the ISR photon
and thus exploits the full information of the event.
We apply this method to an example case of minimal supergravity
(mSUGRA)~\cite{Chamseddine:1982jx,Cremmer:1982vy,mSUGRA}
in which the neutralino is the lightest supersymmetric particle (LSP).
The mSUGRA parameters have been chosen so that the LSP becomes bino-like;
under this assumption, the dominant contribution to the neutralino pair production
becomes the $t$-channel diagram shown in Figure~\ref{fig:tchannel}.
We perform a likelihood analysis based on the photon distributions
and derive cross-section limits.
We assume an integrated luminosity of 500~fb$^{-1}$ at $\sqrt{s}=500$~GeV.
\section{Signal and background samples}
Signal events have been generated using WHIZARD~\cite{whizard}.
A GEANT4-based~\cite{Geant4} simulation of the detector response
was carried out by Mokka~\cite{ilcsoft}
using the International Large Detector (ILD) Concept~\cite{ild-loi}.
For the background studies, we used the full
Standard Model samples which were generated for
the ILD Letter of Intent~\cite{ild-loi}.

Signal samples were generated with two different LSP masses
by varying the mSUGRA parameter $M_{1/2}$, resulting in
LSP masses of approximately 150 and 200~GeV.
The other mSUGRA parameters have been fixed to the following values:
$m_0=300$~GeV, $A_0=0$, $\tan\beta=5$ and $\mathrm{sgn}(\mu)=+1$.
The mass spectra have been computed using SOFTSUSY~\cite{softsusy};
the relevant values are summarized in Table~\ref{tab:ne1ne1a-masses-params}.
\begin{table}[hbt]
\centering
\def\ne{{\tilde\chi^0_1}}
\def\sel{{\tilde e_L}}
\def\ser{{\tilde e_R}}
\begin{tabular}{l|cc}\hline\hline
& Sample A & Sample B \\ \hline
$m_\ne$ (LSP) & 151 GeV & 205 GeV \\
$m_\sel$ & 395 GeV & 453 GeV \\
$m_\ser$ & 333 GeV  & 354 GeV \\ \hline
$M_{1/2}$ & 375 GeV & 500 GeV \\ \hline\hline
\end{tabular}
\caption{Mass spectra of signal samples.}
\label{tab:ne1ne1a-masses-params}
\end{table}
The default cuts of WHIZARD have been modified to allow
the generation of photons with energy greater than 0.1~GeV 
and $|\cos\theta|<0.995$ where the angle $\theta$ is taken to be between
the direction of the photon momentum and the beam axis.
The signal samples have been generated separately for different
polarizations for the electron and positron beams.
For the case of $m_{\tilde\chi_1^0}=150$~GeV, the cross sections
restricted to the energy and angle ranges used for the event generation
are
$\sigma(e^-_R e^+_L\rightarrow\gamma \tilde\chi^0_1\tilde\chi^0_1)=56$~fb
and
$\sigma(e^-_L e^+_R\rightarrow\gamma \tilde\chi^0_1\tilde\chi^0_1)=1.9$~fb;
for the case of 200 GeV, the respective cross sections are 18~fb and 0.55~fb.
The difference in the cross section between the two polarization combinations arises
primarily due to the weak hypercharge of the chiral fields
and, to a lesser extent, the difference in the scalar electron masses.
The two samples were mixed with appropriate weights
to yield a sample with the beam polarizations
$(P_{e^-},P_{e^+})=(+0.8,-0.3)$.

We list the dominant background processes
after the event selection in Table~\ref{tab:dominant-backgrounds}.
\begin{table}[hbt]
\centering
\begin{tabular}{c|l}\hline\hline
Process & Cross section\\ \hline
$\gamma\gamma\rightarrow \ell^+\ell^-$ & $1.1\times10^3$ fb \\
$e^+e^-\rightarrow\gamma\nu_\ell\overline{\nu}_\ell$  & $7.8\times10^2$ fb \\ \hline\hline
\end{tabular}
\caption{Dominant background processes:
the cross sections in the table are
calculated after the event selection as described in Section~\ref{sec:event-selection}.
The lepton generations $(\ell=e,\mu,\tau)$ are summed in the cross section numbers.
}
\label{tab:dominant-backgrounds}
\end{table}

\section{Event reconstruction}
\label{sec:event-selection}
\begin{figure}[hbt!p]
\centering
\includegraphics[width=0.48\columnwidth]{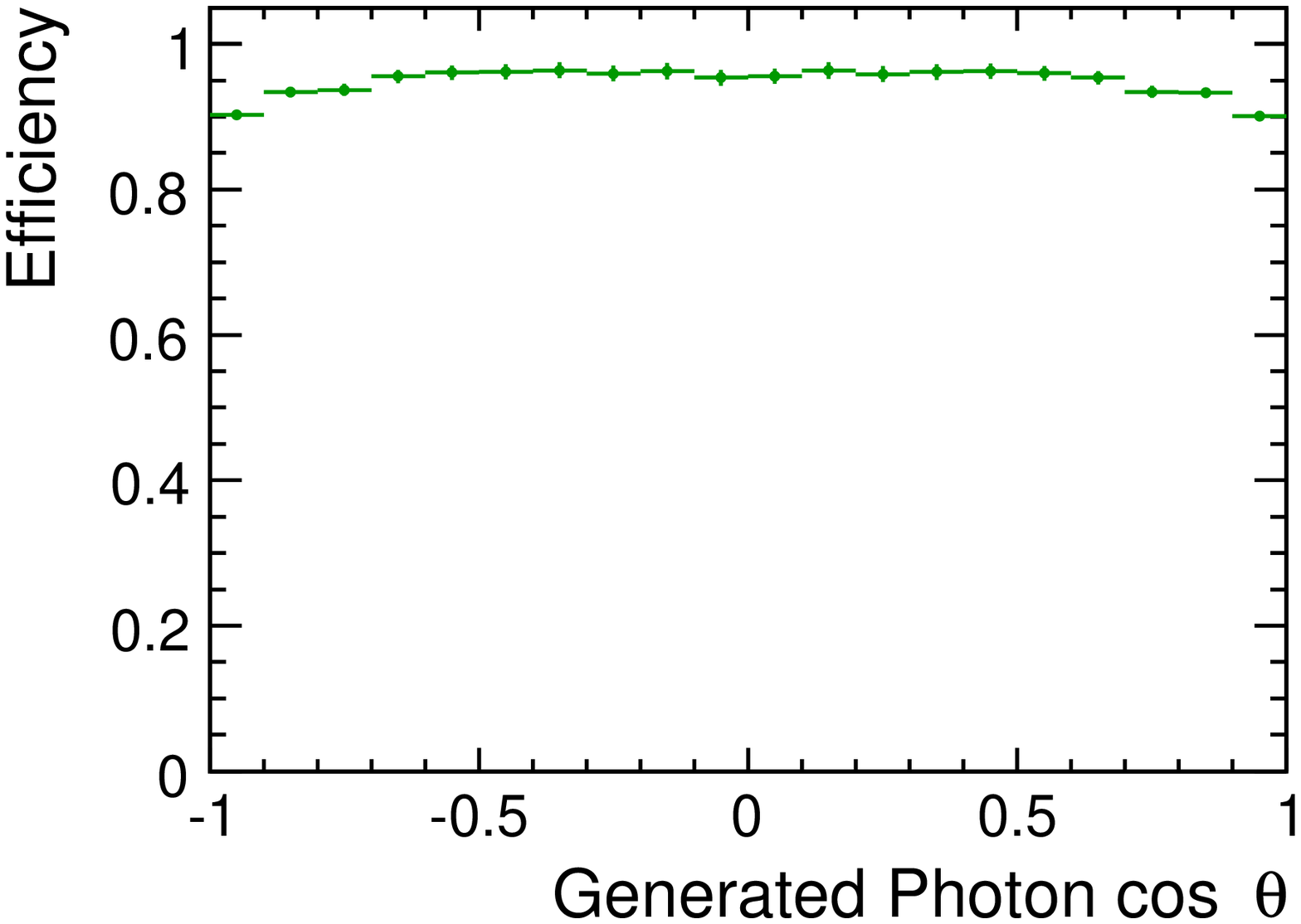}
\includegraphics[width=0.48\columnwidth]{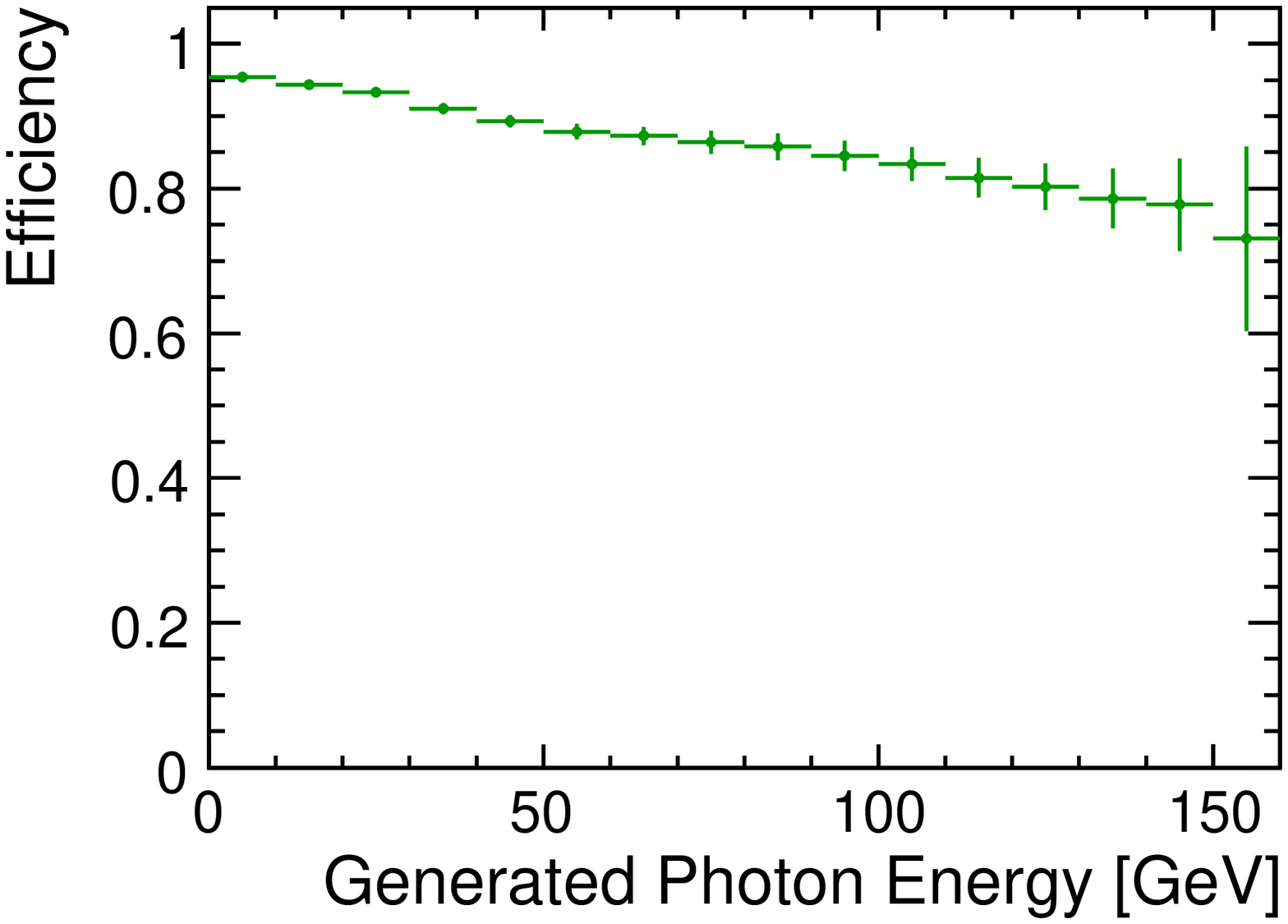}
\caption{Reconstruction efficiency as a function of angle and energy
after the photon cluster merging.}
\label{fig:ne1ne1a-eff}
\end{figure}

We rely on the Pandora Particle Flow Algorithm~\cite{pandora}
for the reconstruction of calorimeter clusters and charged particles.
A single highly energetic photon often results into multiple clusters.
In order to recover the split clusters,
we merge the clusters which lie within a cone of an opening angle of
1.5 degree around the most energetic cluster.
For the event selection, we require that there are no charged particles in the event.
In addition, we require that there are no other clusters aside from the merged cluster.
The merged cluster is required to have an energy greater than 0.5~GeV.
For the particle identification,
we take the energy deposits in the
electromagnetic calorimeter $E_\mathrm{ECAL}$ and the
hadronic calorimeter $E_\mathcal{HCAL}$
to compute the energy fraction
$E_\mathrm{ECAL}/(E_\mathrm{ECAL}+E_\mathrm{HCAL})$
which is required to be greater than 0.9.
The reconstruction efficiency of signal events
as a function of energy and angle
is shown in Figure \ref{fig:ne1ne1a-eff}.
The overall signal efficiency is 94\%
after the event selection.

\section{Likelihood analysis}
\begin{figure}[hbt!p]
\centering
\includegraphics[width=0.32\columnwidth]{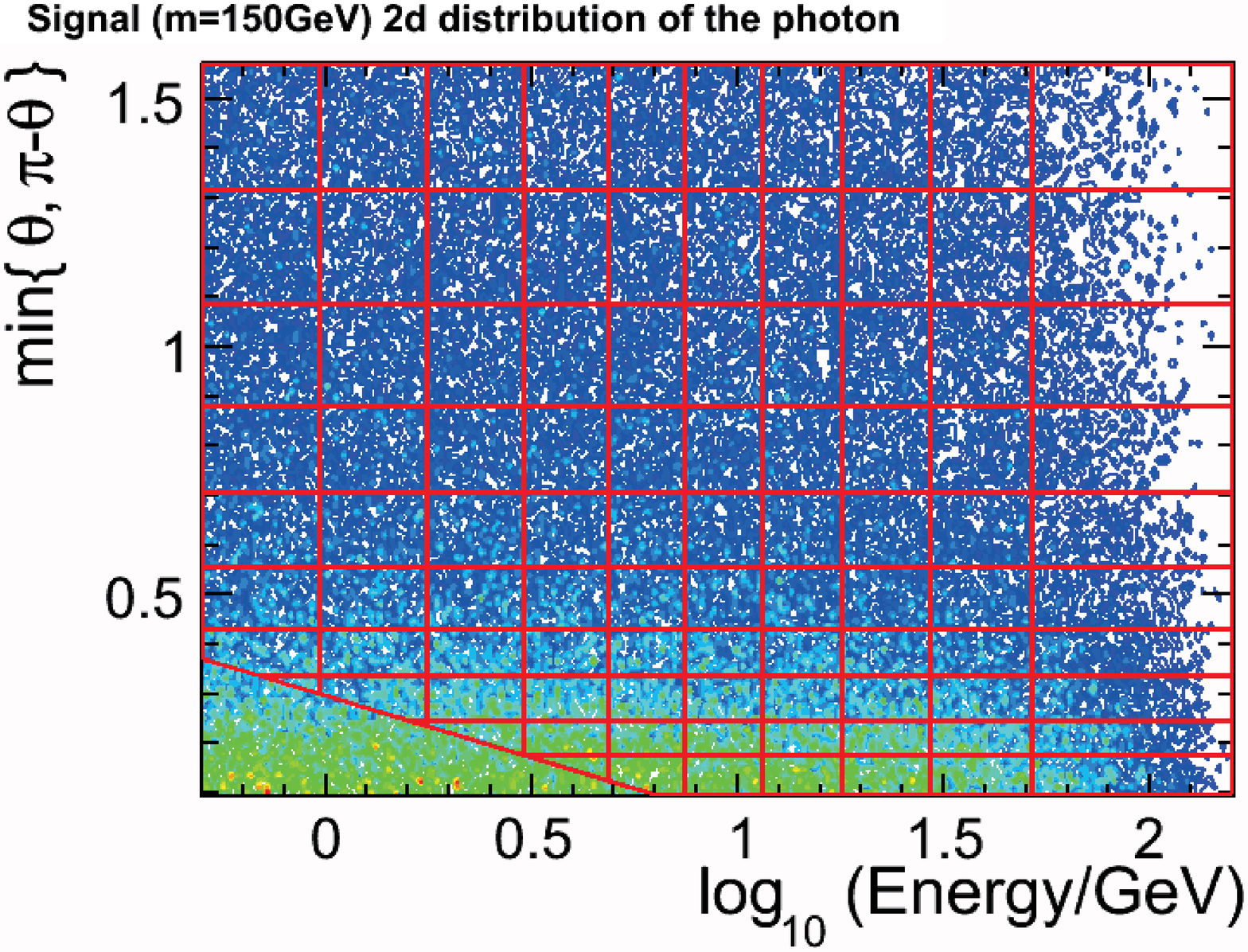}
\includegraphics[width=0.32\columnwidth]{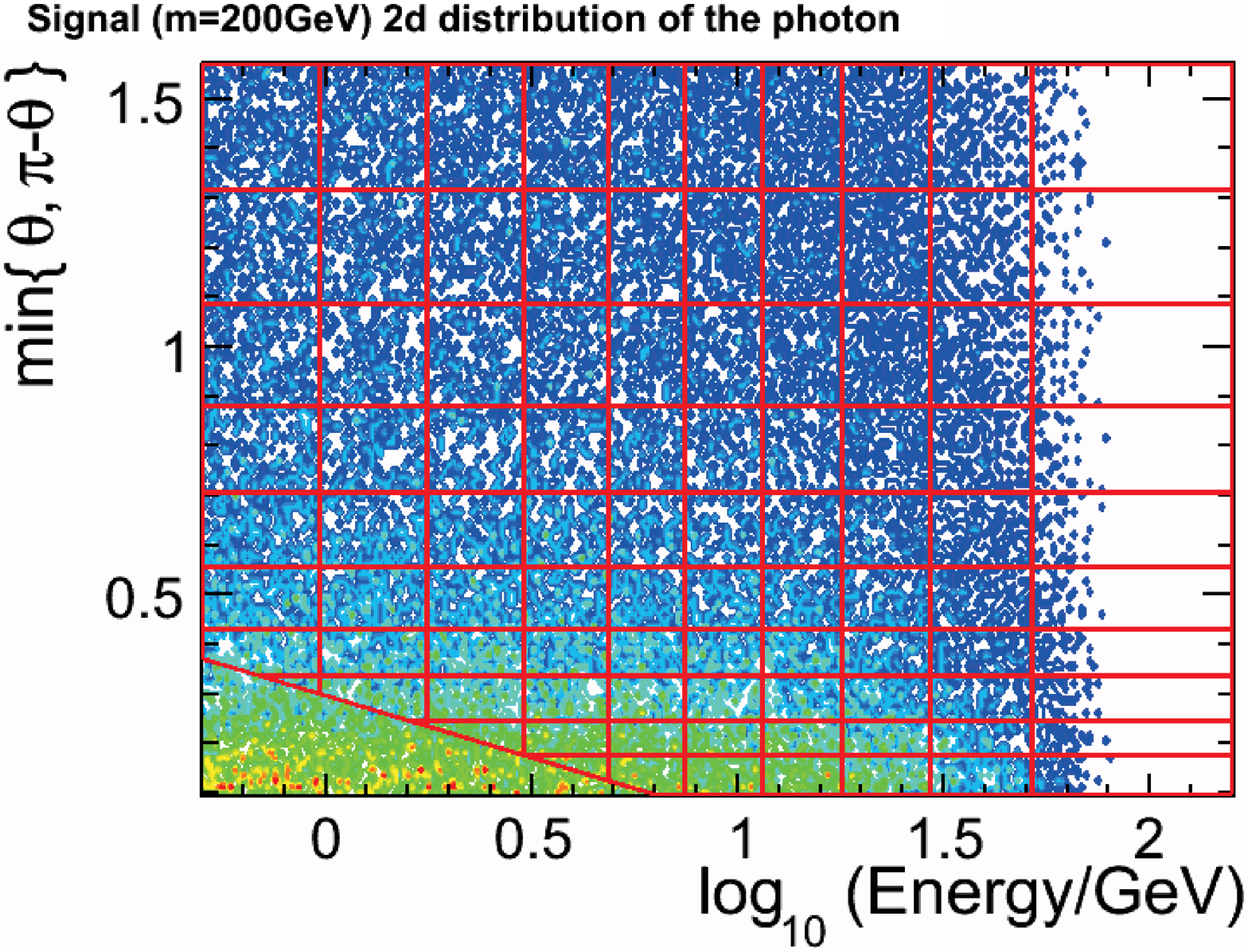}
\includegraphics[width=0.32\columnwidth]{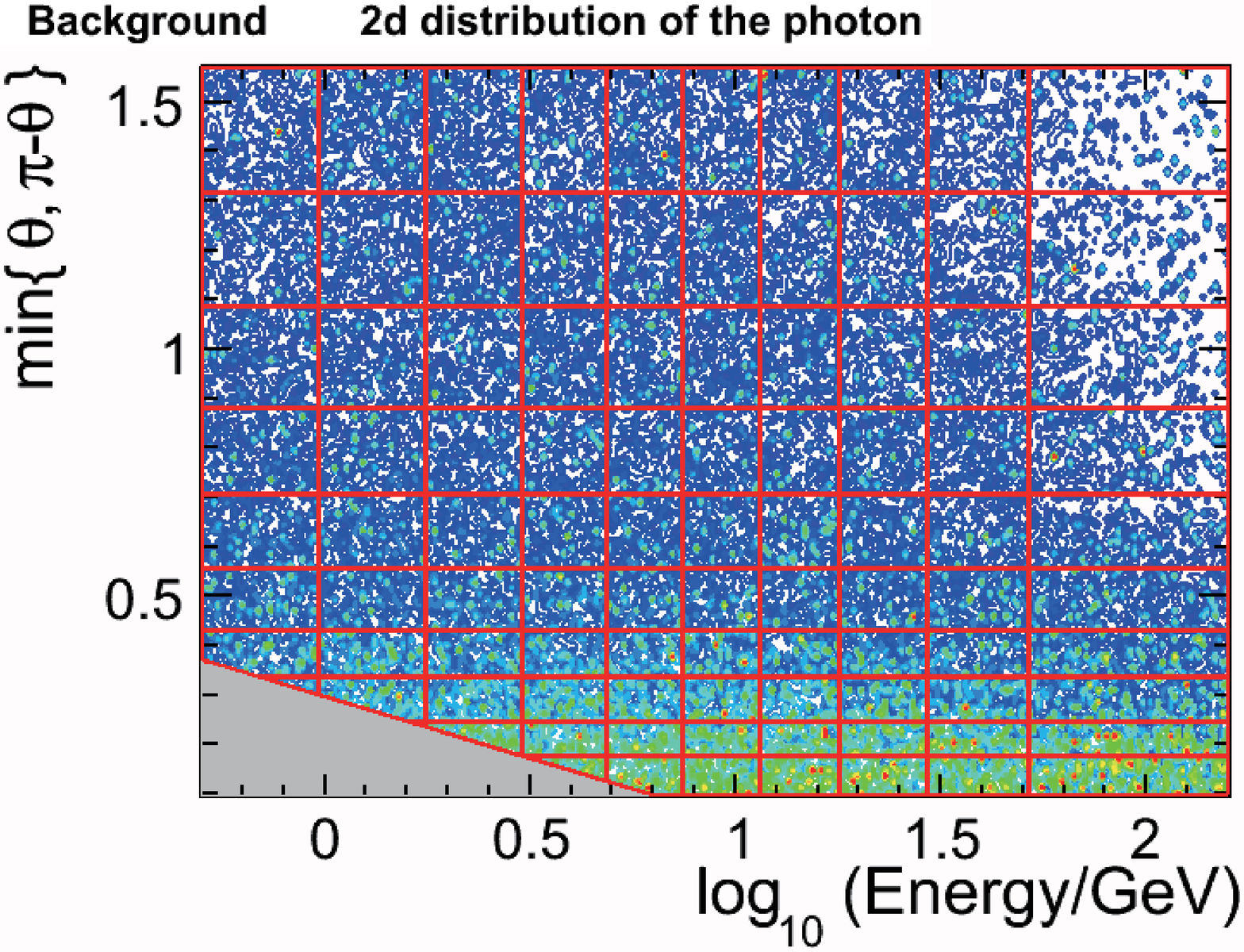}
\caption{
  Two-dimensional distributions for signal events
    with $m_{\tilde\chi_1^0}=150$~GeV (left),
    $m_{\tilde\chi_1^0}=200$~GeV (center),
    and background events (right).
  The horizontal axis is the photon energy in logarithmic scale.
  The vertical axis is the photon angle in radians, folded around $\pi/2$.
  The binning is shown as red lines;
  the lower-left corner is cut off to make the binning
  consistent with the cuts in the background sample.
}
\label{fig:binning}
\end{figure}

\begin{figure}[hbt!p]
\centering
\includegraphics[width=0.60\columnwidth]{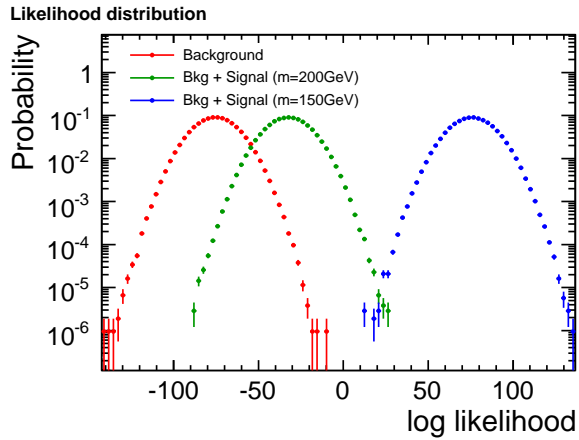}
\caption{Distribution of log-likelihood values for
  background only (red), background and signal with $m_{\tilde\chi_1^0}=200$~GeV (green),
  and background and signal with $m_{\tilde\chi_1^0}=150$~GeV (blue).
  The vertical axis is the probability density for a given log-likelihood value.
}
\label{fig:ne1ne1a-likelidist}
\end{figure}

We perform a binned likelihood analysis using the two dimensional
distribution of the photon energy (logarithmic scale) and angle.
We choose a binning such that the number of bins is $N=96$,
with the binning widths shown in Figure~\ref{fig:binning}.
The choice of $\log E$ instead of $E$ is motivated by the fact that
the ISR photon distribution diverges at low energy.
The angle is folded around $\theta=\pi/2$,
to exploit the symmetry of the distribution.
The bin size is adjusted to ensure the presence of
sufficient signal events in each bin.
The log likelihood ratio is defined as follows:
\begin{equation}
\ln\mathcal{L}(n_1, n_2, \ldots, n_N)
= \sum_{i=1}^{N} \ln\frac{P(n_i;b_i+s_i)}{P(n_i;b_i)}
\end{equation}
where
$n_i$ is the observed number of events in the $i$-th bin,
$b_i$ is the expected number of background events in the $i$-th bin,
$s_i$ is the expected number of signal events in the $i$-th bin,
and
$P(n;\lambda)=\lambda^{n}e^{-\lambda}/n!$ is the
Poisson probability density function
with the expectation value of $\lambda$.
The $m_{\tilde\chi_1^0}=150$~GeV sample was used
to build the template signal shape;
the same signal shape was used to build
likelihood functions even for different masses
in the subsequent analysis, since the true mass is not known {\it a priori}.
The likelihood distributions are obtained by performing
Monte-Carlo test experiments whereby
the observed number of events in each bin is fluctuated
using Poissonian random numbers.
The resulting likelihood distributions for
background plus signal events ($\lambda_i=b_i+s_i$)
are compared with that for
background events only ($\lambda_i=b_i$)
to compute the probability for observing the WIMP signal,
 as shown in Figure~\ref{fig:ne1ne1a-likelidist}.
From this distribution, we calculate the expected
significance for each neutralino mass.
We estimate the significance by computing the probability
that the backgrounds fluctuate to deviate more than
the median of the signal likelihood.
In the case of $m_{\tilde\chi_1^0}=200$~GeV,
the estimated significance is 3.1$\sigma$;
in the case of $m_{\tilde\chi_1^0}=150$~GeV,
the estimated significance is well above 5$\sigma$.
The difference in the significance between the two cases
is mostly due to the difference in the cross section.

In the case of a more general WIMP scenario,
the dependence of the signal cross section
on the WIMP mass is expected to be considerably different.
In order to take this effect into account,
we performed a scan of likelihood values by varying
the WIMP mass and the cross section independently.
The best signal separation power can be attained
when the correct signal template function is used;
assuming this is the case, we keep our mSUGRA signal template.
Additional signal samples were generated using 15 different WIMP mass points
ranging from 100~GeV to 240~GeV by varying $M_{1/2}$.
For this study, we use the beam polarizations of
$(P_{e^-},P_{e^+})=(+1.0,-1.0)$.
We obtain the cross section limit,
defined as the cross section which gives a 50\% probability of
5$\sigma$ observation, as a function of the WIMP mass.
\begin{figure}[hbt!p]
\centering
\includegraphics[width=0.60\linewidth]{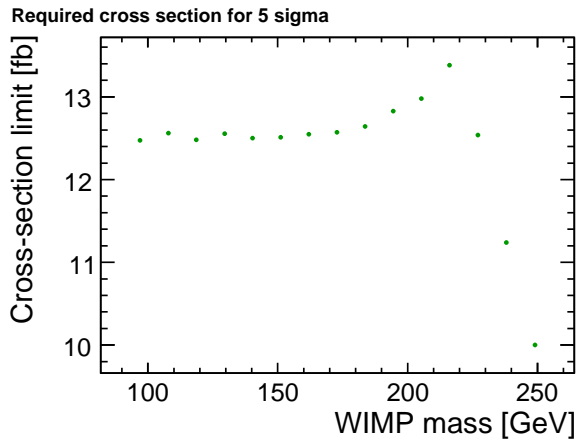}
\caption{
  Cross section limits for WIMP observation.
  The vertical axis is
  the cross-section limit of the polarization combination
  $(P_{e^-},P_{e^+})=(+1.0,-1.0)$
  integrated over the kinematically allowed region in this analysis.
}
\label{fig:ne1ne1a-mscan}
\end{figure}

For a given mass, we adjusted the cross section
until it yielded the desired probability of observing the signal.
We performed $10^6$ Monte-Carlo test experiments
at each step of testing the cross section.
A binary search was used to find the cross section limit efficiently.
The cross section limit as a function of the WIMP mass is plotted in Figure~\ref{fig:ne1ne1a-mscan}.
If the cross section is at least 14~fb, the WIMP can be observed.
Around the mass of 200~GeV, the required cross section increases
since the likelihood function is constructed using the template signal shape for
$m_{\tilde\chi_1^0}=150$~GeV.
As the WIMP mass approaches 250~GeV, the required cross section becomes smaller
since the photon energy distribution peaks at the low energy region,
thereby allowing it to be discriminated against the background shape.
Note, however, that if one fixes the model and let the mass approach 250 GeV,
the cross section decreases faster than the cross-section limit.
Thus, this result is consistent with the fact that the
heavier neutralino is more difficult to observe.

\section{Conclusions}
We have looked at the mSUGRA neutralino search as an example of a WIMP search with single photons
using the two-dimensional distribution of the ISR photon energy and the angle.
We compared the likelihood distributions and calculated the probability to achieve 5$\sigma$ observation.
In our mSUGRA model, the $m_{\tilde\chi_1^0}=150$~GeV case can be easily observed,
while the $m_{\tilde\chi_1^0}=200$~GeV case has the significance of 3.1$\sigma$.
We obtained cross-section limits as a function of the neutralino mass.
WIMPs created through the $t$-channel with the cross section of over 15~fb
can be observed with the 5$\sigma$ significance at the ILC with $\sqrt s=500$ GeV.
In principle, this method can be applied to a broader class of WIMP searches with ISR photons.

\section*{Acknowledgments}
The authors would like to thank all members of the ILC physics subgroup~\cite{ilcwg} for useful discussions.
This study is supported in part by the Creative Scientific Research Grant No.~18GS0202
of the Japan Society for Promotion of Science.

\begin{footnotesize}

\end{footnotesize}



\begin{thebibliography}{99}
\bibitem{Presentation}
Presentation:
http://ilcagenda.linearcollider.org/contributionDisplay.py?contribId=218\&sessionId=17\&confId=4175
\bibitem{Lopez:1996ey}
J.~L.~Lopez, D.~V.~Nanopoulos and A.~Zichichi,
Phys.\ Rev.\  D {\bf 55}, 5813 (1997)
[arXiv:hep-ph/9611437].
\bibitem{Akrawy:1990gy}
  M.~Z.~Akrawy {\it et al.}  [OPAL Collaboration],
  Phys.\ Lett.\  B {\bf 248}, 211 (1990).
\bibitem{Fayet}
P.~Fayet,
Phys.\ Lett.\  B {\bf 117}, 460 (1982);
P.~Fayet,
  Phys.\ Lett.\  B {\bf 175}, 471 (1986).
\bibitem{Fargion:1995qb}
D.~Fargion, M.~Y.~Khlopov, R.~V.~Konoplich and R.~Mignani,
  Phys.\ Rev.\  D {\bf 54}, 4684 (1996).
\bibitem{Birkedal:2004xn}
  A.~Birkedal, K.~Matchev and M.~Perelstein,
      Phys.\ Rev.\  D {\bf 70}, 077701 (2004)
  [arXiv:hep-ph/0403004].
\bibitem{Konar:2009ae}
  P.~Konar, K.~Kong, K.~T.~Matchev and M.~Perelstein,
      New J.\ Phys.\  {\bf 11}, 105004 (2009)
  [arXiv:0902.2000 [hep-ph]].
\bibitem{Bartels}
  C.~Bartels and J.~List,
  {\it In the Proceedings of 2007 International Linear Collider Workshop (LCWS07 and ILC07), Hamburg, Germany, 30 May - 3 Jun 2007, pp COS02}
  [arXiv:0709.2629 [hep-ex]];
  C.~Bartels and J.~List,
  arXiv:0901.4890 [hep-ex].
\bibitem{Chamseddine:1982jx}
  A.~H.~Chamseddine, R.~L.~Arnowitt and P.~Nath,
  Phys.\ Rev.\ Lett.\  {\bf 49}, 970 (1982).
\bibitem{Cremmer:1982vy}
  E.~Cremmer, P.~Fayet and L.~Girardello,
      Phys.\ Lett.\  B {\bf 122}, 41 (1983).
\bibitem{mSUGRA}
  R.~Barbieri, S.~Ferrara and C.~A.~Savoy,
  Phys.\ Lett.\  B {\bf 119}, 343 (1982);
  L.~J.~Hall, J.~D.~Lykken and S.~Weinberg,
  Phys.\ Rev.\  D {\bf 27}, 2359 (1983);
  P.~Nath, R.~L.~Arnowitt and A.~H.~Chamseddine,
  Nucl.\ Phys.\  B {\bf 227}, 121 (1983).
\bibitem{whizard}
  W.~Kilian, T.~Ohl and J.~Reuter,
  arXiv:0708.4233 [hep-ph];
  M.~Moretti, T.~Ohl and J.~Reuter,
  arXiv:hep-ph/0102195.
\bibitem{Geant4}
  S.~Agostinelli {\it et al.}  [GEANT4 Collaboration],
  Nucl.\ Instrum.\ Meth.\  A {\bf 506}, 250 (2003).
\bibitem{ilcsoft}
ILC software portal, http://ilcsoft.desy.de
\bibitem{ild-loi}
  DESY 2009/87, Fermilab PUB-09-682-E, KEK Report 2009-6,
  {\em International Large Detector - Letter of Intent},
  ISSN 0418-9833, ISBN 978-3-935702-42-3.
\bibitem{softsusy}
  B.~C.~Allanach,
  Comput.\ Phys.\ Commun.\  {\bf 143}, 305 (2002)
  [arXiv:hep-ph/0104145].
\bibitem{pandora}
  M.~A.~Thomson,
  Nucl.\ Instrum.\ Meth.\  A {\bf 611}, 25 (2009)
  [arXiv:0907.3577 [physics.ins-det]].
\bibitem{ilcwg}
ILC Physics Subgroup: http://www-jlc.kek.jp/subg/physics/ilcphys/
\end{thebibliography}
\end{document}